\documentclass[12pt]{article}
\usepackage{amssymb}
\usepackage{amsbsy}
\usepackage{epsfig}
\usepackage{verbatim}
\makeatletter \@addtoreset{figure}{section}
\def\thefigure{\thesection.\@arabic\c@figure}
\def\fps@figure{h, t}
\@addtoreset{table}{bsection}
\def\thetable{\thesection.\@arabic\c@table}
\def\fps@table{h, t}
\@addtoreset{equation}{section}

\newtheorem{corollary}{Corollary}[section]
\newtheorem{definition}{Definition}[section]
\newtheorem{theorem}{Theorem}[section]
\newtheorem{proposition}{Proposition}[section] 
\newtheorem{example}{Example}[section]

\newtheorem{lemma}{Lemma}[section]
\newtheorem{remark}{Remark}[section]
\newtheorem{remarks}[remark]{Remarks}

\def\bd{\begin{definition}}
\def\ed{\end{definition}}
\def\bt{\begin{theorem}}
\def\et{\end{theorem}}
\def\bp{\begin{proposition}\rm}
\def\ep{\end{proposition}}
\def\bc{\begin{corollary}}
\def\ec{\end{corollary}}
\def\bl{\begin{lemma}\em}
\def\el{\end{lemma}}
\def\be{\begin{equation}}
\def\ee{\end{equation}}
\def\br{\begin{remark}\rm\small}
\def\er{\end{remark}}
\def\brs{\begin{remarks}.\\ \rm\
\begin{enumerate}}
\def\ers{\end{enumerate}\end{remarks}}
\def\bea{\begin{eqnarray}}
\def\eea{\end{eqnarray}}

\def\ra{{\rightarrow}}

\def \ss {\subset}

\def\tr{\mathrm {tr}}
\def\det{\mathrm {det}}

\def\span{\mathrm {span}}

\def\diag{\mathrm {diag}}
\def\span{\mathrm {span}}
\def\Fr{\mathrm {Fr}}
\def\Gr{\mathrm {Gr}}

\def\&{&{\hskip -20pt}}

\def\BB{\mathcal{B}}

\def\FF {\mathcal{F}}
\def\HH{\mathcal{H}}

\def\MM{\mathcal{M}}

\def\PP{\mathcal{P}}

\def\Ib{\mathbf{I}}

\def\Nb{\mathbf{N}}
\def\Pb{\mathbf{P}}
\def\Rb{\mathbf{R}}

\def\Zb{\mathbf{Z}}

\def\Hb{\mathbf{H}}

\def\grP{\mathfrak{P}}

 \def\grgl{\mathfrak{gl}}

\def\nchi{\hbox{\raise 2.5pt\hbox{$\chi$}}}
\date{}
\begin{document}
\baselineskip 16pt 
\medskip
\begin{center}
\begin{Large}\fontfamily{cmss}
\fontsize{17pt}{27pt}
\selectfont
\textbf{ Convolution symmetries of integrable hierarchies, \break matrix models and $\tau$-functions}\footnote{Work of J.H. supported by the Natural Sciences and Engineering Research Council of Canada (NSERC) and the Fonds Qu\'ebecois de la recherche sur la nature et les technologies (FQRNT).  Work of A.O. supported by RFBR grant  
11-01-00440-a and RAS Program ``Fundamental Methods in Nonlinear Physics''.}
\end{Large}\\
\bigskip
\begin{large}  {J. Harnad}$^{1,2}$
 and {A. Yu. Orlov}$^{3}$
 \end{large}
\\
\bigskip
\begin{small}
$^{1}${\em Centre de recherches math\'ematiques,
Universit\'e de Montr\'eal\\ C.~P.~6128, succ. centre ville, Montr\'eal,
Qu\'ebec, Canada H3C 3J7 \\ e-mail: harnad@crm.umontreal.ca} \\
\smallskip
$^{2}${\em Department of Mathematics and
Statistics, Concordia University\\ 1455 de Maisonneuve Blvd. W. 
Montreal, Quebec,  Canada H3G 1M8} \\ 
\smallskip
$^{3}${\em Nonlinear Wave Processes Laboratory, \\
Institute of Oceanology, 36 Nakhimovskii Prospect, 
Moscow 117851, Russia\\
 e-mail: orlovs55@mail.ru, orlovs@ocean.ru } \\
\end{small}
\end{center}
\bigskip

\begin{center}{\bf Abstract}
\end{center}
\smallskip

\begin{small}

Generalized convolution symmetries of integrable hierarchies of  KP and 2KP-Toda  type act diagonally on the Hilbert space $\HH= L^2(S^1)$ in the standard monomial basis. The induced transformations on the  Hilbert space Grassmannian  $\Gr_{\HH_+}(\HH)$ may be viewed as symmetries of these hierarchies, acting upon the Sato-Segal-Wilson $\tau$-functions, and thereby generating new solutions of the hierarchies. The  corresponding transformations of the associated fermionic Fock space are also diagonal in the standard orthonormal basis,  labeled by integer partitions.  The Pl\"ucker coordinates of the image under the Pl\"ucker map of the element $W \in \Gr_{\HH_+}(\HH)$ defining the initial point under the commuting KP flows are the coefficients in the single and double Schur function expansions of the associated $\tau$ functions. These are therefore multiplied by the eigenvalues of the convolution action in the fermionic representation. Applying such transformations to standard matrix model integrals, we obtain new matrix models of externally coupled type whose partition functions are thus also  seen to be KP or 2KP-Toda  $\tau$-functions. More general  multiple integral representations of tau functions are similarly obtained,  as well as finite determinantal expressions for them. 
 \bigskip
\end{small}
\bigskip \bigskip

\section{Introduction: convolution symmetries of $\tau$-functions }

Solutions of integrable hierarchies of KP and 2KP-Toda type are determined by their $\tau$-functions \cite{Sa, SS, SW}. Infinite sequences of such KP $\tau$-functions $\{\tau(N, {\bf t})\}_{N\in \Zb}$, depending on  the infinite set of commuting flow parameters ${\bf t} = (t_1, t_2, \dots)$ and an integer lattice label $N$,  may be associated in a standard fashion \cite{Sa, SS, SW},  to elements of a ``universal phase'' space, viewed as an infinite Grassmann manifold or flag manifold. These  satisfy the Hirota bilinear equations of the KP hierarchy and also, in certain cases (e.g. exponential flows of matrix model integrals induced by trace invariants),  the equations of the Toda lattice hierarchy. 

The $\tau$-functions may be expanded as infinite series in a basis of Schur functions $s_\lambda ({\bf t})$,  labelled by integer partitions 
$\lambda = (\lambda_1 \ge \lambda_2  \ge \cdots \ge 0)$
\be
\tau(N, {\bf t}) = \sum_{\lambda} \pi_N(\lambda) s_\lambda ( {\bf t}).
\ee
In the approach of Sato and Segal-Wilson \cite{Sa, SW}, the coefficients $ \pi_N(\lambda)$ are interpreted as  
Pl\"ucker coordinates of the image $\PP(W)$ of an element $W$ of a Hilbert space Grassmannian  $Gr_{\HH_+}(\HH)$ under the Pl\"ucker map  
\be
\PP: Gr_{\HH_+}(\HH) \ra \Pb (\FF) 
\ee
into the projectivation of the semi-infinite exterior space $\FF :=\Lambda \HH$ (the Fermionic Fock space). In \cite{SW}, the Hilbert pace $\HH$ is  chosen as the square integrable functions  
$L^2(S^1)$ on the unit circle in the complex $z$-plane and  the elements of $Gr_{\HH_+}(\HH)$ are subspaces of $\HH=L^2(S^1)$ that are ``commensurable'' with the subspace $\HH_+ \ss \HH$  of functions  admitting a holomorphic extension to the interior disk.

The image $\PP(Gr_{\HH_+}(\HH))$ of the Grassmannian under the Pl\"ucker map consists of all decomposable elements of $\Lambda \HH$, which is the intersection of the infinite set of quadrics defined by the Pl\"ucker relations. The latter are equivalent to the infinite set of Hirota bilinear differential relations  \cite{JM, Sa, SS} for  $\tau(N, {\bf t})$, which are the defining property of $\tau$-functions. Through the Sato formula for the Baker-Akhiezer function
\be
\Psi_N (z, {\bf t}) =  e^{\sum_{i=1}^\infty t_i z^i} {\tau(N, {\bf t} - [z^{-1}]) \over \tau(N, {\bf t} )}, \qquad [z^{-1}] := (z^{-1}, 2z^{-2}, 3 z^{-3},  \dots ),
\label{SatoBA}
\ee
these equations are equivalent to the KP hierarchy and their associated Lax equations.

The 2KP-Toda hierarchy \cite{JM, UT} can similarly be expressed in terms of  $\tau$-functions 
depending on $N$,  ${\bf t}$ and a further infinite sequence of flow parameters
${\bf \tilde t} = (\tilde{t}_1, \tilde{t}_2, \dots)$. These admit double Schur function expansions \cite{T}
\be
\tau^{(2)}(N, {\bf t},  {\bf \tilde t}) = 
\sum_{\lambda} \sum_{\mu} B_N(\lambda, \mu) s_\lambda( {\bf t}) s_\mu( {\bf \tilde t}),
\ee
in which the coefficients $B_N(\lambda, \mu) $ have a similar interpretation in terms of  Pl\"ucker coordinates. They also satisfy an infinite set of bilinear differential Hirota-type  relations in both sequences of flow variables and difference-differential equations relating different lattice points. For fixed $N$, they include the KP Hirota equtions of the KP hierarchy in each of the two sets of flow variables, so we refer to them as 2KP-Toda tau functions.

Starting with any given $\tau$-function of KP-Toda or 2KP-Toda type,  it will be shown in the following that  new $\tau$-functions can be constructed, satisfying the same sets of bilinear relations, having the following Schur function expansions:
\bea
\tilde{C}_{\rho}(\tau)(N, {\bf t}) &\&= \sum_{\lambda} r_\lambda(N) \pi_N(\lambda) s_\lambda ( {\bf t})
\label{convsymm1} \\
\tilde{C}^{(2)}_{\rho, \tilde{\rho}}(\tau^{(2)})(N, {\bf t},  {\bf \tilde t}) &\& = 
\sum_{\lambda} \sum_{\mu}  r_\lambda(N) B_N(\lambda, \mu)  \tilde{r}_\mu(N) s_\lambda( {\bf t}) s_\mu( {\bf \tilde t}),
\label{convsymm2}
\eea
where the factors $ r_\lambda(N)$, $\tilde{r}_\lambda(N)$ are defined in terms of a given pair of infinite sequences of non-vanishing constants $\{r_i\}_{i\in \Zb}$, $\{\tilde{r}_i\}_{i\in \Zb}$ through the formulae
\be
r_\lambda(N):=c_r (N) \prod_{(i,j) \in \lambda} r_{N-i+j}, \quad 
\tilde{r}_\mu(N):= c_{\tilde{r}} (N)\prod_{(k, l) \in \mu }\tilde{r}_{N-k+l}.
\ee
Here the products are over pairs of positive integers $(i,j) \in \lambda$ and  $(k,l) \in \mu$ that lie within the matrix locations represented by the Young diagrams of the partitions $\lambda$ and $\mu$, respectively,
\be
c_r (N) :=\prod_{i=1}^\infty {\rho_{N-i} \over \rho_{-i} },
\ee
and
\be
r_i = {\rho_i \over \rho_{i-1}} .
\ee
The sequence of nonvanishing parameters $\{\rho_i\}$ may be viewed as Fourier coefficients  of a function $\rho(z)$ on the unit circle, or a distribution.
It will be shown ({\bf Proposition \ref{convtauschurexp}}) that,  in terms of the  elements of the subspace  $W \subset L^2(S^1)$ corresponding to a point of the Grassmannian,  the transformations  (\ref{convsymm1}), (\ref{convsymm2}) mean taking a generalized convolution product with  $\rho(z)$ (and similarly for $\tilde{r}_i$). These will therefore be referred to as (generalized)  {\it convolution symmetries}. 

With the usual  2KP-Toda flow parameters $({\bf t}, \tilde{\bf t})$  fixed at some specific  values, such transformations  extend to an infinite  abelian group of commuting flows whose parameters determine the $\rho_i$'s. This has been used to generate new classes of solutions of integrable hierarchies \cite{BAW, O2, OSc}.  In the present work, they are studied rather as individual transformations, for fixed  values  of the parameters $\rho_i$ which, when applied to a given KP-Toda or 2KP-Toda $\tau$-function, produce a new one. Particular cases that implicitly use such transformations as symmetries have found applications, e.g.,  as generating functions for topological invariants related to Riemann surfaces, such as Gromov-Witten invariants and Hurwitz numbers \cite{Ok, OkP}.

As an immediate application, we may start with  an integral over $N\times N$ Hermitian matrices:
\be
Z_N({\bf t}) = \int_{M\in \Hb^{N \times N}} d\mu(M) \,    e^{\tr\sum_{i=1}^\infty t_i  M^i}, 
\ee
where $d\mu$ is a suitably defined $U(N)$ conjugation invariant measure on the space 
$ \Hb^{N\times N}$ of Hermitian $N \times N$ matrices,. This is known to be a KP-Toda $\tau$-function 
  \cite{ZKMMO}.  Applying a convolution symmetry (\ref{convsymm1}) with $\rho(z)$ taken  essentially as the exponential function $e^z$ on the unit disc, and evaluating at flow parameter values 
\be
t_i ={1\over i}  \tr (A^i), \quad {\bf t}= [A] = (t_1, t_2, \dots )
\label{tiA}
\ee
for a fixed $N \times N$ Hermitian  matrix  $A$ we obtain, within a constant multiplicative factor, the externally coupled matrix model integral ({\bf Proposition \ref{convdefZNext_exp}}).
\be
Z_{N, ext} ([A]) := \int _{M\in \Hb^{N\times N}} d\mu(M) e^{\tr A M} = (\prod_{i=1}^{N-1}i! )\, \tilde{C}_{\rho}(Z_N)([A]). 
\ee
Such integrals arise in a number contexts, such as  the Kontsevich-Witten generating function \cite{K}, the Brezin-Hikami model  \cite{BH, ZJ1, ZJ2} and the complex Wishart ensemble \cite{BS, DW}. More general choices for the function $\rho(z)$ are shown in {\bf Proposition  \ref{convdefZNext}} to also determine  KP-Toda $\tau$-functions as externally coupled matrix  integrals. It is further shown, 
in {\bf Proposition  \ref{deterpZNrhoA}},  that these matrix model $\tau$-functions can be expressed as  finite $N \times N$ determinants. 

Similarly, Hermitian two-matrix integrals with exponential coupling of Itzykson-Zuber type \cite{IZ}
\be
Z^{(2)}_N({\bf t},  {\bf \tilde t})
= \int_{M_1 \in  \Hb^{N\times  N}} d\mu(M_1)\int_{M_2  \in \Hb^{N\times  N}}  d\tilde{\mu}(M_2) \ e^{\tr(\sum_{i=1}^\infty( t_i M_1^i +\tilde{ t}_i M_2^i)+ M_1M_2)} 
\label{IZ2integral}
\ee
are known to be 2KP-Toda $\tau$-functions \cite{AvM1, HO1, HO2, O1}. Applying the convolution symmetry (\ref{convsymm2}) to  (\ref{IZ2integral})  gives   an externally coupled two-matrix integral 
({\bf Proposition~\ref{convZN2extAB}}).
\be
 \tilde{C}^{(2)}_{\rho, \tilde{\rho}}(Z_N^{(2)})([A], [B]) = \int_{M_1 \in  \Hb^{N\times  N}} \hspace{-24 pt}d\mu(M_2)\int_{M_2  \in \Hb^{N\times  N}}  \hspace{-24 pt} d\tilde{\mu}(M_2)\  \tau_r (N, [A], [M_2])  \tau_{\tilde{r }}(N, [B], [M_2])
e^{\tr( M_1M_2)},
\ee
where $[A]$ and  $[B]$   signify the sequences $\{{1\over i}\tr(A^i)\}_{i\in \Nb^+}$ and 
$\{{1\over i}\tr(B^i)\}_{i\in \Nb^+}$ of trace invariants for the pair of Hermitian matrices $A$ and $B$ and 
\bea
\tau_r (N, [A], [M_1]) &\&= \sum_\lambda  r_\lambda(N)s_\lambda ([A]) s_\lambda ([M_1]),
\\
\tau_{\tilde{r}} (N, [B], [M_2]) &\&= \sum_\lambda  \tilde{r}_\lambda(N)s_\lambda ([B]) s_\lambda ([M_2]).
\eea
This doubly externally coupled two-matrix model $\tau$-function can also be expressed in a finite $N\times N$ determinantal form (eq.~(\ref{CauchyBinet_SchurABM2}), {\bf Proposition \ref{andreif_rho_tilderho}}). 

This approach can also be extended to more general 2KP-Toda  $\tau$-functions admitting multiple integral representations of the form (\ref{2KPmult_int_tau}). Applying the convolution symmetry 
(\ref{convsymm2}) then gives a new  2KP-Toda  $\tau$-function expressible either as a multiple integral (eq.~(\ref{conv2KP_multiple_int}), {\bf Proposition \ref{conv2KP}}) or as a finite determinant 
 (eq.~(\ref{CauchyBinet_SchurABM3}), {\bf Proposition \ref{conv_multiple_int_det}}). 

 The key to understanding these constructions, and further results following from them, is the interpretation of the Sato $\tau$-function as a vacuum state expectation value of  products of exponentials of bilinear combinations of fermionic creation and annihilation operators  \cite{Sa, JM, UT}. This well-known construction will be summarized in the next section. 
 
\section{Fermionic construction of $\tau$-functions}

   We recall here the approach  to the construction of $\tau$-functions for integrable hierarchies of the KP and Toda types due to Sato \cite{Sa, SS}, the Kyoto school \cite{DJKM1, DJKM2, JM, UT} and Segal and Wilson \cite{SW}.

\subsection {Hilbert space Grassmannian and fermionic Fock space }

  We begin with the ``first quantized'' Hilbert space $\HH$, which will be identified, as in \cite{SW}, with the space of square integrable functions on the unit circle
   \be
   \HH = L^2(S^1) = \HH_+ + \HH_-,
\ee
decomposed as the direct sum of the subspaces  $\HH_+=\span\{z^i\}_{i\in \Nb}$ and  $\HH_- =\span\{z^{-i}\}_{i\in \Nb^+} $ consisting of functions that admit holomorphic extensions, respectively, to the interior and exterior of the unit circle $S^1$ in the complex $z$-plane, with the latter vanishing at $z=\infty$. For consistency with other conventions, the monomial (orthonormal) basis elements $\{e_i \}_{i\in \Zb}$ of $\HH$ will be denoted  
\be
e_i := z^{-i-1},  \quad i \in \Zb.
\label{eibasis}
\ee
   
     Two infinite abelian groups act on $\HH$ by multiplication: 
 \be \Gamma_+:=\{\gamma_+({\bf t}):=e^{\sum_{i=1}^\infty t_i z^i}\},
 \quad {\rm and \quad}  \Gamma_- :=\{\gamma_-({\bf t}):=e^{\sum_{i=1}^\infty t_i z^{-i}}\},
 \ee
  where  ${\bf t}:= (t_1, t_2, \dots)$  is an infinite sequence of (complex) flow parameters corresponding to the one-parameter subgroups.   More generally, we have the general linear group $GL(\HH)$ consisting of invertible endomorphisms connected to the identity with well defined determinants. (See \cite{SW} for more detailed definitions of this and what follows.) 

        We consider the Grassmannian $\Gr_{\HH_+}(\HH)$ of subspaces $W \subset \HH$ that are {\it commensurable} with $\HH_+ \subset \HH$ (in the sense of \cite{SW}; i.e.,  that orthogonal projection $\pi_+: W \ra \HH_+$ to $\HH_+$  is a Fredholm operator while  projection $\pi_-:W \ra \HH_-$ to $\HH_-$  is Hilbert-Schmidt). The connected components of $\Gr_{\HH_+}(\HH)$, denoted   $\Gr_{\HH^N_+}(\HH)$, $N\in \Zb$, consist of those $W \in \Gr_{\HH_+}(\HH)$ for which the Fredholm index of  $\pi_+:W \ra \HH_+$ is $N$.  These are the $GL(\HH)$ orbits of the subspaces 
\be
   \HH_+^N := z^{-N}\HH_+  \subset  \HH,
   \ee
     whose elements are denoted $W_{g,N} = g( \HH_+^N)  \in \Gr_{\HH^N_+}(\HH)$.  The solutions to the KP hierarchy are given by the $\tau$-function $\tau_{N, g} ({\bf t})$ as defined below,  which determines the orbit of $W_{g,N} $ in $\Gr_{\HH^N_+}(\HH)$ under $\Gamma_+$  through its Pl\"ucker coordinates. In the terminology \cite{SW}, the index $N$ is called the ``virtual dimension'' of the elements $W_{g,N} \in \Gr_{\HH^N_+}(\HH)$; i.e. their dimension {\it relative} to the those in the component $\Gr_{\HH^0_+}(\HH)$ containing $\HH_+$.
            
         The {\it Fermionic Fock space} is the exterior space $\FF:=\Lambda \HH$ consisting of (a completion of) the span of the semi-infinite wedge products:
     \be
  \vert \lambda, N\rangle := e_{l_1} \wedge e_{l_2} \wedge \cdots ,
 \ee
  where $\{l_j\}_{j\in \Nb^+}$ is a strictly decreasing sequence of integers that saturates, for sufficiently large $j$, to a descending sequence of consecutive integers. This is equivalent to requiring that there be an associated  pair $(\lambda, N)$ consisting of  an integer $N$  and a partition $\lambda = (\lambda_1, \dots,  \lambda_{\ell(\lambda)}, 0 ,  0, \dots)$  of length $l(\lambda)$ and weight  $\vert \lambda\vert =\sum_{i=1}^{\ell(\lambda)} \lambda_i$,  where the parts $\lambda_i$ are a weakly decreasing  sequence of non-negative integers that are positive for $i \le \ell(\lambda)$, and zero for $i > \ell(\lambda)$,  such that the sequence $\{l_j\}_{j\in \Nb^+}$ is given by
    \be
  l_j := \lambda_j - j +N. 
\ee
  In particular, for the trivial partition $\lambda = (0)$, we have the ``charge $N$ vacuum" vector
   \be
  \vert 0, N\rangle = e_{N-1} \wedge e_{N-2} \wedge \dots ,
\ee
 which will henceforth be denoted  $\vert  N\rangle$.   The full Fock space $\FF$ thus admits a decomposition 
as an  orthogonal direct sum of the subspaces $\FF_N$  of states with charge $N$
\be
\FF = \bigoplus_{N\in \Zb} \FF_N.
\ee

 Denoting by $\{\tilde{e}^i\}_{i \in \Zb}$ the basis for $\HH^*$ dual to the monomial basis $\{e_i\}_{i\in \Zb}$ for $\HH$, we define the Fermi creation and annihilation operators $\psi_i$ and $\psi^\dag_i$ on an arbitrary vector $v \in \FF$ by exterior and interior multiplication, respectively:
   \be
 \psi_i v = e_i \wedge v,  \quad \psi^\dag_i v := i_{\tilde{e}^i} v, \quad v\in \FF.
 \label{psi_psidag}
\ee
These satisfy the standard canonical anti-commutation relations generating the Clifford algebra on $\HH + \HH^*$ with respect to the natural corresponding quadratic form
\be
[\psi_i, \psi_j]_+ = [\psi^\dag_i, \psi^\dag_j]_+ =0, \quad [\psi_i, \psi^\dag_j]_+ = \delta_{ij}.
\ee
The basis states $\vert \lambda, N\rangle$  may be expressed in terms of  creation and annihilation operators acting upon the charge $N$ vacuum vector as follows \cite{HO4} 
 \be
  \vert \lambda, N\rangle  = (-1)^{\sum_{i=1}^k \beta_i}\prod_{i=1}^k \psi_{N+\alpha_i} \psi^\dag_{N-\beta_i -1} \vert N\rangle , 
     \ee
where $(\alpha_1, \dots \alpha_k \vert  \beta_1, \dots , \beta_k)$ is the Frobenius notation (see
 \cite{Mac}) for the partition $\lambda$; i.e.,  $\alpha_i$ is the number of boxes in the corresponding Young diagram to the right of the $i$th diagonal element and $\beta_i$ the number below it.
     
  The Pl\"ucker map $\grP : \Gr_{\HH_+}(\HH) \ra \Pb(\FF)$ takes the subspace 
  $W = \span (w_1, w_2, \dots )$ into the projectivization of the exterior product of its basis elements:
    \be
  \grP: \span (w_1, w_2, \dots ) \mapsto \left[ w_1 \wedge w_2 \wedge  \cdots  \right], 
 \ee
 and may be lifted to a map from the bundle $\Fr_{\HH_+}(\HH)$ of frames on  $\Gr_{\HH_+}(\HH)$
  to $\FF$ 
   \bea
  \hat{\grP} : \Fr_{\HH_+}(\HH) \quad  &\&\ra \quad  \FF \cr
 \hat{ \grP}:  (w_1, w_2, \dots )&\& \mapsto w_1 \wedge w_2 \wedge  \cdots  .
  \eea
These  interlace the lift of the action of the abelian group $\Gamma_+ \times \HH \ra \HH$ to $\Fr_{\HH_+}(\HH)$ or $\Gr_{\HH_+}(\HH)$ with the following representation of  $\Gamma_+$ on $\FF$ (and its projectivization)
   \be
\gamma_+({\bf t}): v \mapsto  \hat{\gamma}_+({\bf t}) v, \quad   \hat{ \gamma}_+({\bf t}) : = e^{\sum_{i=1}^\infty t_i H_i},   \quad v\in \FF,
\ee
where
   \be
H_i := \sum_{n\in \Zb} \psi_n \psi^\dag_{n+i},  \quad i \in \Zb , \ i\ne 0,
\ee
and ${\bf t} = (t_1,  t_2,  \dots )$ is the infinite sequence of flow parameters.
Similarly, the Pl\"ucker maps  $\hat{\grP}$ and $\grP$   interlace  the lift of the action of the abelian group $\Gamma_- \times \HH \ra \HH$ to $\Fr_{\HH_+}(\HH)$ or $\Gr_{\HH_+}(\HH)$ with the following representation  of  $\Gamma_-$ on $\FF$ (and its projectivization).
   \be
\gamma_-({\bf t}): v \mapsto  \hat{\gamma}_-({\bf t}) v, \quad    \hat{\gamma}_-({\bf t})
 : = e^{\sum_{i=1}^\infty t_i H_{-i}},  \quad v\in \FF, 
\ee

   \br
   Note that the image under the Pl\"ucker map of the virtual dimension $N$ component $\Gr_{\HH_+^N}(\HH)$  of the Grassmannian $\Gr_{\HH_+}(\HH)$ is the $GL(\HH)$ orbit of the charged vacuum state  $|N\rangle$, consisting of all decomposable elements of $\FF_N$. 
     \er
     
    The KP-Toda $\tau$-function $\tau_g({N, \bf t})$ corresponding to the element
$W_{g,N}  \in \Gr_{\HH_+}(\HH)$ is given, within a nonzero multiplicative constant, by applying the group elements  $\gamma_+({\bf t})$  to $W_{g,N} $,  to obtain the $\Gamma_+$ orbit $\{W_{g,N}({\bf t}) := \gamma_+({\bf t}) (W_{g,N})\}$, and taking the linear coordinate (within projectivization) of the image under the Pl\"ucker map   corresponding to projection along the basis element $\vert N\rangle$
       \be
    \tau_g (N, {\bf t}) = \langle  N \vert  \hat{\grP}(W_{g,N}({\bf t}))\rangle.
  \ee
    If the group element $g\in GL(\HH) $ is interpreted, relative to the monomial basis $\{e_i\}_{i\in \Zb}$, as an infinite matrix exponential  $g=e^A$ of an element of the Lie algebra $A \in \grgl(\HH)$ with matrix elements $A_{ij}$, then  the corresponding representation of $GL(\HH)$ on $\FF$ is given by
   \be
    \hat{g} := e^{\sum_{i,j \in \Zb} A_{ij} : \psi_i \psi^\dag_j : } ,
\ee
where $:\  :$ denotes normal ordering (i.e. annihilation operators $\psi_j^\dag$ appearing to the right
when $j\ge 0$ and creation operators $\psi_i$ to the right when $i < 0$).
This gives the following expression for $\tau_{N,g}({\bf t})$ as a charge $N$ vacuum state
    expectation value of a product of exponentiated bilinears in the Fermi creation and
    annihilation operators
   \be
    \tau_g(N,{\bf t})= \langle N \vert  \hat{\gamma}_+ ({\bf t})\hat{g} \vert N\rangle.
    \label{tauNg}
 \ee  
   The equations of the KP  hierarchy are then equivalent to the well-known infinite system of Hirota bilinear equations \cite{JM, Sa, SS} which, in turn, are just the Pl\"ucker relations for the decomposable element ${\grP}(W_{g, N}({\bf t})) \in \Pb(\FF_N)$.
    
Similarly, we may define a $2$-Toda sequence of  double KP $\tau$-functions associated to the group element $\hat {g}$ 
   \be
\tau^{(2)}_g(N, {\bf t}, \tilde{\bf t}) =    \langle N \vert  \hat{\gamma}_+({\bf t})\hat{g}\hat{\gamma}_-(\tilde{\bf t}) \vert N\rangle, 
\label{tau2Ng}
\ee
where $\tilde{\bf t} = (\tilde{t}_1,  \tilde{t}_2,  \dots )$ is a second infinite set of flow parameters. This may similarly be shown to satisfy the Hirota bilinear relations of the 2KP-Toda hierarchy.

\subsection {Schur function expansions}

Evaluating the matrix elements of $\hat{\gamma}_+({\bf t})$ and $\hat{\gamma}_-({\bf t})$ between
the states $\vert N\rangle$ and $\vert \lambda, N\rangle$ gives  the Schur function
\be
\langle N \vert  \hat{\gamma}_+({\bf t})\vert \lambda, N\rangle = \langle \lambda, N\vert  \hat{\gamma}_-({\bf t})\vert  N\rangle  =s_\lambda({\bf t}), 
\ee
(cf. \cite{Sa, SS, HO2, HO3} which  is determined  through the Jacobi-Trudy formula
\be
s_\lambda({\bf t}) = \det(h_{\lambda_i -i +j}({\bf t}))\vert _{1\le i,j \le \ell(\lambda)}
\ee
in terms of the complete symmetric functions $h_i({\bf t})$, defined by
\be
e^{\sum_{i=1}^\infty t_i z^i} = \sum_{i=0}^\infty h_i({\bf t}) z^i.
\ee

Inserting a sum over a complete set of intermediate states in eqs.~(\ref{tauNg}), (\ref{tau2Ng}), we obtain  the single and double Schur functions expansions 
\bea
  \tau_ g(N, {\bf t}) &\&= \sum_\lambda \pi_{N,g}(\lambda) s_\lambda ({\bf t}),  
  \label{tauNgschur}
  \\
\tau^{(2)}_g (N, {\bf t}, \tilde{{\bf t}}) &\&= \sum_\lambda \sum_\mu B_{N,g}(\lambda, \mu) s_\lambda ({\bf t})
s_\mu (\tilde{{\bf t}}).
  \label{tau2Ngschur}
\eea
Here the sum is over all partitions $\lambda$ and $\mu$ and
\be
 \pi_{N,g}(\lambda)  =  \langle \lambda, N \vert  \hat{g} \vert  N\rangle
 \label{pluckerdef1}
 \ee
 is the Pl\"ucker coordinate  of  the image of the element   $g(\HH_+^N)  \in \Gr_{\HH_+^N}(\HH)$ under the Pl\"ucker map $\grP$ along the basis direction $\vert \lambda, N\rangle$  in the charge $N$  sector  $\FF_N$ of the Fock space. Similarly,
 \be
B_{N,g}(\lambda, \mu)  =  \langle \lambda, N \vert   \hat{g} \vert  \mu, N\rangle
 \label{pluckerdef2}
 \ee
  may be viewed as the $\vert \lambda, N\rangle$ Pl\"ucker coordinate of the image of the element $g(w_{\mu, N}) \in \Gr_{\HH_+^N}(\HH)$, where
    \be
  w_{\mu, N} := \span\{e_{\mu_i -i +N} \} \in \Gr_{\HH_+^N}(\HH).
  \ee
  
  In particular, choosing $g$ to be the identity element $\Ib $, and using Wick's theorem (or equivalently, the Cauchy-Binet identity in semi-infinite form), we obtain \cite{HO2}
   \be
\tau^{(2)}_\Ib (N, {\bf t}, \tilde{\bf t}) =   \langle N \vert  \hat{\gamma}_+({\bf t})\hat{\gamma}_-(\tilde{\bf t}) \vert N\rangle
= \sum_\lambda s_\lambda ({\bf t}) s_\lambda (\tilde{{\bf t}})
=e^{\sum_{i=1}^\infty i t_i \tilde{t}_i},
\label{tau2NId}
\ee
where the last equality is the Cauchy-Littlewood identity (cf. ref.~\cite{Mac}).

\section{Convolution symmetries}
\subsection {Convolution action on $\HH$ and $\Gr_{\HH_+}(\HH)$}

Consider now an infinite sequence of  complex numbers $\{T_i\}_{i\in \Zb}$, 
and define
\be
\rho_i := e^{T_i}, \quad \quad  r_i := {\rho_i \over \rho_{i-1}},  \quad i \in \Zb.
\label{rhoidef_ridef}
\ee 
In the following, we will assume that the series $\sum_{i=1}^\infty T_{-i}$ converges and that
\be
\lim_{i\ra \infty}| r_i| = r \le 1,
\label{rlim}
\ee
(although, for some purposes, the latter condition may be weakened).
It follows that the two series
\be
\rho_+(z)  =\sum_{i=0}^\infty \rho_i z^i, \quad \rho_-(z) = \sum_{i=1}^\infty \rho_{-i}  z^{-i}, 
\label{rhodef}
\ee
are absolutely convergent  in the interior and exterior  of the unit circle \hbox{$|z|=1$}, respectively, defining analytic functions $\rho_\pm(z)$ in these regions and that
\be
 R_\rho := \prod_{i=1}^\infty \rho_{-i}
\label{rhoprod}
\ee
converges to a finite value.  If the inequality (\ref{rlim}) is  strict, $\rho_+(z)$ extends to  the unit circle, defining  a function in  $L^2(S^1)$.  Henceforth, we denote the pair $(\rho_+, \rho_-)$ by $\rho$,
where the latter can be viewed as a sum $\rho_- + \rho_+$ in the sense of distributional convolutions, as defined below.

   If $w \in L^2(S^1)$ has the Fourier series decomposition
   \be
 w(z)  = \sum_{i=-\infty}^\infty w_i e_i = \sum_{i=-\infty}^\infty w_i z^{-i-1} = w_+(z) + w_-(z)
  \ee
 where
 \be
 w_+(z) := \sum_{i=0}^\infty w_{-i-1} z^{i}, \quad w_-(z):= \sum_{i=1}^\infty w_i z^{-i-1},
 \label{wFourier}
   \ee
   (note the different labelling conventions in (\ref{rhodef}) and (\ref{wFourier})),
 we can define a bounded linear map $C_\rho: L^2(S^1) \ra L^2(S^1)$ that has the effect of multiplying each Fourier coefficient $w_i$ by the factor $\rho_{i}$, and hence each basis element $e_i$ by $\rho_i$:
   \be 
   C_\rho(w)(z) =  \sum_{i=-\infty}^\infty \rho_i w_i z^{-i-1} = C_\rho(w)_+ +   C_\rho(w)_-
   \label{convmap}
   \ee
   This can be interpreted as taking a convolution product with the  function (or distribution)
   \be
  \tilde\rho(z) = \tilde{\rho}_+(z) + \tilde{\rho}_-(z) 
      \label{tilderho}
    \ee
    where
    \bea
     \tilde{\rho}_+(z)  &\&:= z^{-1} \rho_-(z^{-1})  =\sum_{i=0}^\infty \rho_{-i-1} z^{i}, \\
      \tilde{\rho}_-(z)  &\&:= z^{-1} \rho_+(z^{-1} ) =\sum_{i=0}^\infty \rho_i z^{-i-1}, 
     \label{tilderhoplusminus}
        \eea
    
    \bea
    C_{\rho}(w)_+(z) &\&:= \lim_{\epsilon \ra 0^+} {1\over 2\pi i } \oint_{|\zeta| = 1-\epsilon}
       \tilde{\rho}_+({\zeta }) w_+(z/\zeta)\zeta^{-1} d\zeta , 
        \label{conv+}\\
         C_{\rho}(w)_-(z) &\& := \lim_{\epsilon \ra 0^+} {1\over 2\pi i } \oint_{|\zeta |= 1+\epsilon}
       \tilde{ \rho}_-(\zeta) w_-(z/\zeta) \zeta^{-1}d\zeta
         \label{conv-}        \eea
     (with the contour integrals taken counterclockwise).
   
    If  $\rho_-(z)$ extends analytically to $S^1$,  eq.~(\ref{conv+}) is
    an ordinary convolution product on the circle (in exponential variables).  In the examples detailed below,  all but a finite number of the  $T_{-i}$ values vanish for $i > 0$, and hence the infinite
    product  (\ref{rhoprod}) is really finite, but $\rho_-(z)$ is rational with a pole at $z=1$ and the convolution product  (\ref{conv-}) may be understood on $S^1$ only in the sense of distributions.
  
\br
Note  that the class of generalized convolution mappings defined by (\ref{convmap}) - (\ref{conv-}) only forms a semi-group since, although they may be invertible, their inverse does not  generally belong to the same class. It may be extended to a group by dropping the condition (\ref{rlim}), or restricted to one by requiring $r=1$, but this will not be needed in the sequel. The linear maps $C_\rho:\HH \ra \HH$ may nevertheless be interpreted as elements of   $GL(\HH)$, and are simply represented in the monomial basis $\{e_i\}$ by the diagonal matrix   $\diag\{\rho_i\}$. They thus belong to the abelian subgroup of $GL(\HH)$ consisting of invertible elements  that are diagonal in the monomial basis. 
 \er
 
 \br
Since the Baker-Akhiezer function (\ref{SatoBA}), evaluated at  all values of the parameters
${\bf t} = (t_1, t_2, \dots )$, spans the shifted element $z^{N} (W_{g, N}) \in \Gr_{\HH_+^0}(\HH)$ in the zero virtual dimension component of the Grassmannian, the convolution action (\ref{conv+}), 
(\ref{conv-}),  lifted to the Grassmannian, may be obtained by applying its conjugate $z^{N} \circ  C_\rho \circ z^{-N}$ under the shift map $z^{-N}: \Gr_{\HH_+^0}(\HH) \ra \Gr_{\HH_+^N}(\HH)$  to 
$\Psi_N(z, {\bf t})$. But note that, at fixed values of the flow parameters ${\bf t}$, this does not equal the value of the Baker-Akhiezer function corresponding to the transformed $\tau$-function as defined below; only the subspaces of $\HH$ that they span,  varying over the ${\bf t}$ values,  will coincide. This fact will not be used explicitly in the following, but it underlies the geometrical meaning of  generalized convolutions as symmetries of KP-Toda and 2KP-Toda hierarchies.
\er
 
\subsection {Convolution action on Fock space}
We now consider the action  $\hat{C} \times\FF \ra \FF$ of the abelian subgroup  of $GL(\HH)$ consisting of diagonal elements in the monomial basis, and associate an element $\hat{C}_\rho \in \hat{C}$ to each sequence $\{\rho_i\}_{i\in \Zb}$  defined as above, such that the Pl\"ucker map $\hat{\grP}$ intertwines  the $\hat{C}_\rho$ action with that of $C_\rho$,  lifted to the bundle $\Fr_{\HH_+}(\HH)$ of frames over $\Gr_{\HH_+}(\HH)$, and is equivariant with respect to group multiplication in $\hat{C}$.

  To do this, we first introduce the abelian algebra generated by the operators
  \bea                                                                                                                                     
 K_i &\&:= \  : \! \psi_i \psi^\dag_i {\hskip -3 pt}: \ =  
 \  \cases{\phantom{-}  \psi_i \psi^\dag_i  \quad {\rm if} \quad i \ge 0 \cr
   -\psi_i^\dag \psi_i \quad {\rm if}    \quad i < 0 , }
 \label{abelianK}  \\
 {}
[K_i, K_j ]&\& \phantom{:}  = 0,    \quad    i,j \in \Zb.
\eea
For $\{\rho_i = e^{T_i}\}_{i\in \Zb}$  as above, define the operator
\be
\hat{C}_\rho :=e^{\sum_{i=-\infty}^{\infty} T_i K_i}.
\ee
\begin{definition} For each pair $(\lambda, N)$,  where $N\in \Zb$, and $\lambda$ is a partition which, expressed in Frobenius notation, is  $(\alpha_1 \cdots \alpha_k \vert \beta_1 \cdots \beta_k)$, let 
\be r_\lambda(N) := c_r(N) \prod_{(i,j) \in \lambda} r_{N-i+j} = 
c_r(N)  \left(\prod_{i=1}^k {\rho_{N+ \alpha_i} \over \rho_{N-\beta_i -1}}\right),
 \label{rlambdaNdef}
\ee
where
\be
c_r(N) := \cases{ \prod_{i=0}^{N-1}\rho_i  \quad {\rm if} \quad N > 0 \cr
           \quad  \ 1 \qquad \ \     {\rm if} \quad  N=0  \cr
        {1\over \prod _{i=N}^{-1} \rho_i} \quad \ \  {\rm if } \quad N < 0.}
        \label{crNdef}
\ee
Here the inclusion $(i,j) \in \lambda$ is understood to mean that  the matrix location $(i,j)$ corresponds to a box within the Young diagram of the partition $\lambda$; i.e., $1 \le i \le \ell(\lambda)$, $1 \le j \le \lambda_i$. The second equality in
(\ref{rlambdaNdef}) follows from the definition (\ref{rhoidef_ridef}). 
\end{definition}
It follows that $\hat{C}_\rho$ acts diagonally in the basis $\{ \vert \lambda, N \rangle\}$, with
eigenvalues $r_\lambda(N)$.

\begin{lemma}
\be
\hat{C}_\rho |\lambda, N\rangle =  r_\lambda(N)  |\lambda, N\rangle.
\label{diagonalChataction}
\ee
\label{diagChataction}
\end{lemma}
\noindent{\bf Proof}:
Since the Fock space basis element $|\lambda, N\rangle$ is an infinite wedge product 
\bea
|\lambda, N\rangle &\&= e_{l_1}\wedge e_{l_2} \wedge \cdots   =
 (-1)^{\sum_{i=1}^k \beta_i }  
\prod_{i=1}^k\psi_{N+\alpha_i}  \psi^\dag_{N-\beta_i -1} \vert N\rangle,
\label{Nlambda_wedge} \\
l_j &\& := \lambda_j - j + N, \quad j \in \Nb^+ ,
\eea
 it follows from the definition (\ref{psi_psidag}) and the normal ordering in  (\ref{abelianK})  that the effect of the action of  $e^{T_i K_i}$  on $|\lambda, N\rangle$  is to introduce a multiplicative factor $\rho_i$  if  $i\ge 0$ and   $e_i$ is present in the wedge product (\ref{Nlambda_wedge})  or  $\rho_i^{-1}$ if $i < 0$  and it is absent, and otherwise no factor. Therefore
\bea
\hat{C}_\rho |\lambda, N\rangle  &\&= \hat{C}_\rho (-1)^{\sum_{i=1}^k \beta_i }  
\prod_{i=1}^k \psi_{N+\alpha_i} \psi^\dag_{N-\beta_i -1} \vert N\rangle \cr
&\& = {\prod_{i=1}^\infty \rho_{N-i}\over \prod_{i=1}^\infty \rho_{-i}}   \left(\prod_{i=1}^k {\rho_{N+ \alpha_i} \over \rho_{N-\beta_i -1}}\right) \vert \lambda, N \rangle\cr
&\& = c_r(N)  \left(\prod_{i=1}^k {\rho_{N+ \alpha_i} \over \rho_{N-\beta_i -1}}\right) \vert \lambda, N\rangle
\cr
&\& = r_\lambda(N) \vert \lambda, N\rangle.
\eea
   \hfill Q.E.D.

Now let $W = \span\{w_i(z) \in L^2(S^1)\}_{i\in \Nb^+} \in \Gr_{\HH_+}(\HH)$ and view $\{w_i\}_{i\in \Nb^+}$ as a frame for $W$.
\begin{lemma}
 The Pl\"ucker map $\hat{\grP}$ intertwines the convolution action (\ref{convmap}) and the $\hat{C}$-action on $\FF$
\be 
\hat{\grP}(\{C_\rho(w_i)\}_{i\in \Nb^+}) =  R_\rho \hat{C}_\rho (\hat{\grP}\{w_i\}_{i\in \Nb^+}), 
\label{convintertwine}
\ee
with multiplicative factor $R_\rho := \prod_{i=1}^\infty \rho_{-i} $. 
\label{Cintertwine}
\end{lemma}

\noindent{\bf Proof}: Applying $C_\rho$ to each element $w_i \in L^2(S^1)$ defining the frame  for $W\in \Gr_{\HH_+}(\HH)$ just multiplies its Fourier coefficients by the factors $\rho_j$ as in  eq.~(\ref{convmap}). It follows that the basis element $|\lambda, N \rangle$ is multiplied by the product of the factors $\rho_{l_j}$ corresponding to the terms $e_{l_j}$ it contains, as in  (\ref{diagonalChataction}).  Eq.~(\ref{convintertwine}) then follows from the definition of the Pl\"ucker map $\hat{\grP}$ and linearity.  \hfill Q.E.D.

\begin{example}
 \label{example1}
 {\rm \small
Choose
\bea
\rho_+(z) &\&= e^z =  \sum_{i=0}^\infty {z^i \over i!},  \qquad \quad  |z| \le 1 
\label{expconv}
\\
\rho_-(z) &\&= {1 \over z -1} = \sum_{i=1}^\infty z^{-i}    \quad \ \,  |z| >1,
\label{identconv}
\eea
so
\bea
\rho_i  &\&= \cases{ {1\over i!}  \quad {\rm if}  \quad i \ge 1 \cr
                              1  \quad {\rm if} \quad i \le 0, }  \\
r_i  &\&= \cases{{1\over i} \quad {\rm if}  \quad i \ge 1 \cr
  1  \quad {\rm if} \quad i \le 0, } \\
r_\lambda(N) &\&= {1 \over (\prod_{i=1}^{N-1} i!) (N)_\lambda } \quad {\rm if} \quad  \ell(\lambda) \le N
\label{rlambdaN_ex1}
 \eea
 where
 \be
 (N)_\lambda    := \prod_{i=1}^{\ell(\lambda)} \prod_{j=1}^{\lambda_i}(N - i +j)
\ee
is the extended Pochhammer symbol.
}
\end{example}
\begin{example}
 \label{example2}
 {\rm \small
Choose
\be
\rho_+(z) = {1\over (1- \zeta z)^a} =  \sum_{i=0}^\infty (a)_i {(\zeta z)^i \over i!},  \qquad  |\zeta| < 1,\quad
 |z| \le 1 
\ee
and $\rho_-(z)$ again as in (\ref{identconv}), 
so
\bea
\rho_i  &\&= \cases{ {(a)_i {\zeta ^i \over i!}}  \quad \, {\rm if}  \quad i \ge 1 \cr
                            \quad   1 \qquad  \quad  {\rm if} \quad i \le 0, }  \\
r_i  &\&= \cases{{a-1 +i\over i}\zeta \qquad \  {\rm if}  \quad i \ge 1 \cr
\quad  1  \qquad \quad \,  {\rm if} \quad i \le 0, } \\
r_\lambda(N)  &\& = \left( \prod_{i=0}^{N-1}  {(a)_i\over i!}\right)  {\zeta^{|\lambda| +{1\over 2}N(N-1)}(a-1 +N)_\lambda\over (N)_\lambda }     \quad {\rm if} \quad  \ell(\lambda) \le N .
\eea
}
\end{example}

\subsection {Convolutions and Schur function expansions of $\tau$-functions}

We now consider the KP-Toda tau function
\be
  \tau_{C_\rho g}(N, {\bf t} )= \langle N \vert  \hat{\gamma}_+ ({\bf t})\hat{C}_{\rho} \hat{g} \vert N\rangle
  \label{tauNCrhog}
\ee
obtained by replacing the group element $g$ in (\ref{tauNg}) by $C_{\rho} g$. Such a 
$\tau$-function, obtained from  $  \tau_g$ by applying a convolution symmetry will be denoted
\be
  \tau_{C_\rho g}=: \tilde{C}_\rho ( \tau_{g}).
  \label{convsymmKP}
  \ee
Introducing a second pair $(\tilde{\rho}_+, \tilde{\rho}_-)$, 
defined as in (\ref{rhodef}), with the Fourier coefficients $\rho_i$ replaced by $\tilde{\rho}_i$, we also consider the 2-Toda tau function 
   \be
\tau^{(2)}_{C_{\rho}gC_{\tilde{\rho}}} (N, {\bf t}, \tilde{\bf t}) =  \langle N \vert  \hat{\gamma}_+({\bf t})\hat{C}_{\rho}\hat{g}\hat{C}_{\tilde{\rho}}\hat{\gamma}_-(\tilde{\bf t}) \vert N\rangle
\label{tau2NCrhog}
\ee
obtained by replacing  the group element $g$ in (\ref{tau2Ng}) by $C_{\rho} g C_{\tilde{\rho}} $, and denote this transformed 2-Toda $\tau$-function
\be
\tau^{(2)}_{C_{\rho}\hat{g}C_{\tilde{\rho}}} =:
\tilde{C}^{(2)}_{(\rho, \tilde{\rho})}(\tau^{(2)}_g).
  \label{convsymm2KP}
\ee
Inserting sums over complete sets of intermediate orthonormal basis states in (\ref{tauNCrhog}) and 
(\ref{tau2NCrhog}), and defining  ${\tilde{r}}_\lambda(N) $ as in (\ref{rlambdaNdef}), with the factors $\rho_i$ replaced by ${\tilde{\rho}}_i$, we obtain the following form for the Schur function expansions (\ref{tauNgschur}), (\ref{tau2Ngschur}).

\begin{proposition}
\label{conv_tau}
The effect of the convolution actions (\ref{convsymmKP}),  (\ref{convsymm2KP}) is to  multiply the coefficients in the Schur function  expansions of $\tau_{C_\rho g}(N, {\bf t} )$ and $\tau^{(2)}_{C_{\rho}\hat{g}C_{\tilde{\rho}}} (N, {\bf t}, \tilde{\bf t})$ by the diagonal factors $ r_\lambda(N)$ and $\tilde{r}_\mu(N)$.
\bea
  \tau_{C_{\rho}g} (N, {\bf t})&\&= \sum_\lambda r_\lambda(N) \pi_{N,g}(\lambda) s_\lambda ({\bf t}),  
\label{convtauNgschurexp} \\
\tau^{(2)}_{C_{\rho} gC_{\tilde{\rho}} } (N, {\bf t}, \tilde{{\bf t}}) &\&=
\sum_\lambda \sum_\mu r_\lambda(N) B_{N,g}(\lambda, \mu){\tilde{r}}_\mu(N) s_\lambda ({\bf t})
s_\mu (\tilde{{\bf t}}).
\label{convtau2Ngschurexp}
\eea
The Pl\"ucker coordinates  for the modified Grassmannian elements   $C_{\rho}g(\HH_+^N)$ and 
$C_{\rho} gC_{\tilde{\rho}}(w_{\mu, N})$ are thus
\bea
\pi_{N,C_{\rho}g}(\lambda) &\& =   r_{\lambda}(N)  \pi_{N,g}(\lambda) \\
B_{N,C_{\rho} gC_{\tilde{\rho}}}(\lambda, \mu)&\& = 
 r_\lambda(N) B_{N,g}(\lambda, \mu)\tilde{r}_\mu(N). 
\eea
\label{convtauschurexp}
\end{proposition}
\noindent{\bf Proof}: This follows immediately from the diagonal form (\ref{diagonalChataction}) of the $\hat{C}$ action in the orthonormal  basis  $\{|\lambda, N\rangle\}$, substituted into the expansions (\ref{tauNgschur}), (\ref{tau2Ngschur}), using the definitions 
(\ref{pluckerdef1}) and  (\ref{pluckerdef2}) of the Pl\"ucker coordinates $\pi_{N,C_{\rho}g}(\lambda)$ and 
$B_{N,C_{\rho} gC_{\tilde{\rho}}}(\lambda, \mu)$.  \hfill Q.E.D.

In particular, setting $g = C_{\tilde{\rho}} =\Ib$,  in (\ref{convtau2Ngschurexp}) we obtain
\be
\tau^{(2)}_{C_{\rho}} (N, {\bf t}, \tilde{{\bf t}}) = 
\sum_\lambda r_\lambda(N)s_\lambda ({\bf t}) s_\lambda (\tilde{{\bf t}})  =: \tau_r (N, {\bf t}, \tilde{{\bf t}})
\label{hypergeom2KPtau}
\ee
where $\tau_r (N, {\bf t}, \tilde{{\bf t}})$ is defined by the second equality. Such $\tau$-functions have been studied as generalizations of hypergeometric functions in \cite{OSc, O2}. (Cf.  also \cite{HO2, HO3},  where the notation differs slightly  due to the presence of the normalization factor $c_r(N)$
in the definition (\ref{rlambdaNdef}) of $r_\lambda(N)$.)

In the following, the infinite sequence of parameters ${\bf t} =(t_1, t_2, \dots)$ will often be chosen as the trace invariants of some square matrix $M$.
The sequence so formed will be denoted
\be
{\bf t} = [M] = \left\{ {1\over i}\tr(M^i) \right\}\Bigg\vert_{i\in \Nb^+},  \quad [M]_i := {1\over i}\tr(M^i).
\ee
If ${\bf t}$ and  $\bf {\tilde{t}}$  in (\ref{hypergeom2KPtau}) are replaced
by $[A]$ and $[B]$, respectively, where $A$ and $B$ are a pair of diagonal matrices 
\be
A=\diag(a_1, \dots , a_N), \quad 
B=\diag(b_1, \dots , b_N)
\ee
with distinct eigenvalues, and
\be
\Delta(A):= \prod_{1\le i <  j} ^n (a_i - a_j), \qquad
\Delta(B):= \prod_{1\le i <  j} ^n (b_i - b_j)
\ee
denote the Vandermonde determinants in the variables $\{a_i\}$ and $\{b_i\}$,
we obtain a simple  $N\times N$ determinantal expression for $\tau_r (N, [A], [B]) $
(cf.~\cite{HO3, O1}).
\begin{lemma}  Choosing $\rho_-(z)$as in (\ref{identconv}) (i.e. $\rho_{-i}=1$ for $i<1$), we have
\label{dethypergeom}
\bea
\tau_r (N, [A], [B]) &\&= \sum_{\ell(\lambda) \le N} r_\lambda(N)s_\lambda ([A]) s_\lambda ([B])
\label{hypergeom2KPtauAB}\\
&\& =  {\det( \rho_+(a_i b_j)\big\vert_{1\le i,j \le N}  \over \Delta(A)\Delta(B)}.
\label{CauchyBinet_Schur}
\eea
\end{lemma}
\br
Although various proofs of this result may be found elsewhere (e.g., cf.~\cite{HO3}), we provide a detailed version here, based on the Cauchy-Binet identity in semi-infinite form, since it involves some useful further relations. An equivalent way is to use the fermionic form of Wick's theorem, which is really just the Cauchy-Binet identity expressed in terms of  fermionic operators and matrix elements.
\er
\hfill \break \noindent
{\bf Proof of Lemma \ref{dethypergeom}:}
 The Cauchy-Binet identity in semi-infinite form may be expressed by considering two $N$-dimensional framed subspaces $\span \{F_i\}_{1\le i \le N}$ and $\span \{G_i\}_{1\le i \le N}$ of the complex Euclidean vector space $\ell^2(\Nb) =\span\{e_i\}_{i\in \Nb}$, identified with $\HH_+\subset \HH= L^2(S^1)$, by choosing the monomials $\{z^i\}_{i\in \Nb}$ as orthonormal basis. 
 The vectors $F_i$ and $G_j$ are thus identified with elements $F_i(z), G_i(z) \in \HH_+$ defined by
 \be
 F_i(z):= \sum_{j=0}^\infty F_{ji} z^j,  \quad G_i(z):= \sum_{j=0}^\infty G_{ji} z^j.
 \ee
 (Note that, to avoid needless use of negative indices, we are not using the same labelling conventions here for the basis elements $\{e_i\}$ as in (\ref{eibasis}).)
 The complex inner product $(\ , \ )$ is  defined by integration
\be
(F\, , G\,) := {1\over 2\pi i}\oint_{z\in S^1} F(z) G(z^{-1}) {dz \over z}.
\label{innerproduct}
\ee
The Cauchy-Binet identity can then be expressed as
\be
\det(F_i, G_j)\vert_{1\le i,j \le N} = \sum_{\ell(\lambda)\le N}
 \det(F_ {\lambda_i - i +N, j})\det(G_ {\lambda_i - i+N, j}),
  \label{CauchyBinet}
 \ee
where
\be
F_i = \sum_{j\in \Zb} F_{ji} e_j,  \qquad G_i = \sum_{j\in \Zb} G_{ji} e_j,
\ee
and the sum is over all partitions $\lambda$  of length $\ell(\lambda) \le N$, completed so that the $N \times N$ submatrices $F_ {\lambda_i - i +N, j}$ and $G_ {\lambda_i - i +N, j}$ are defined by setting  $\lambda_i =0$ for $i > \ell(\lambda)$. 
Since all expressions in the sum will be polynomials in the parameters $(a_i, b_i)$ there is no loss of generality in assuming that  these lie within the unit disc. We define
\be
F_i(z) := \rho_+(a_i z),  \qquad G_i(z) := (1- b_i z)^{-1}
\ee
and hence
\be
F_{ij} = \rho_i (a_j),  \quad    G_{ij} = (b_j)^i.
\ee
From the character formula
\be
s_\lambda([A]) ={\det(a_i^{\lambda_j-j+N}) \over \Delta(A)}, \quad
s_\lambda([B]) ={\det(b_i^{\lambda_j-j+N}) \over \Delta(B)},
\ee 
it follows that the determinant factors on the RHS of (\ref{CauchyBinet}) are
\bea
 \det(F_ {\lambda_i - i +N, j})  = \det( a_j^{\lambda_i-i+N} \rho_{\lambda_i-i+N})
&\& = \left(\prod_{i=1}^N \rho_{\lambda_i-i+N}\right) s_\lambda([A]) \Delta([A]), \\
 \det(G_ {\lambda_i - i +N, j}) = \det( b_j^{\lambda_i-i+N}) 
&\& =s_\lambda([B]) \Delta(B).
 \eea
 From the definitions (\ref{rlambdaNdef}) and (\ref{crNdef}), it follows that
 \be
  \left(\prod_{i=1}^N \rho_{\lambda_i-i+N}\right) = r_\lambda(N),
 \ee
so the RHS of  the Cauchy-Binet identity (\ref{CauchyBinet}) is just the RHS of 
 eq.~(\ref{hypergeom2KPtauAB})  multiplied by $\Delta([A]) \Delta([B])$.
 On the other hand, from (\ref{innerproduct}), the LHS of (\ref{CauchyBinet}) is 
 \bea
 \det(F_i, G_j) &\&=\det\left( {1\over 2\pi i} \oint_{z\in S^1}  {\rho_+(a_iz)\over z - b_j} {dz \over z} \right)\cr
 &\& \cr
   &\& = \det(\rho_+(a_i b_j)),
 \eea
 which is just the  expression (\ref{CauchyBinet_Schur})  multiplied by $\Delta([A]) \Delta([B])$.
\hfill Q.E.D.
\br 
Note that, for the case of {\bf Example \ref{example1}}, eq.~(\ref{CauchyBinet_Schur}) becomes
the key identity	 (cf. \cite{HO3, ZJ2})
\be
\sum_{\ell(\lambda) \le N} {1\over (N)_{\lambda}}s_\lambda ([A]) s_\lambda ([B])
 = \left(\prod_{k=1}^{N-1} k!\right)  {\det(e^{a_i b_j})\big\vert_{1\le i,j \le N}  \over \Delta(A)\Delta(B)},
\ee
which, together with the character integral  \cite{Mac}
\be
d_{\lambda, N} \int_{U\in U(N)} d\mu_H(U) s_\lambda([AUXU^\dag])  =  s_\lambda([A\,]) s_\lambda([X]),
\label{haar_schur_int}
\ee
(where $d\mu_H(U)$ is the Haar measure on $U(N)$),  implies the Harish-Chandra-Itzykson-Zuber (HCIZ) integral \cite{IZ}
\be
 \int_{U\in U(N)} d\mu_H(U) e^{\tr(AUXU^\dag)}  = 
 \left(\prod_{k=1}^{N-1} k!\right)  {\det(e^{a_i x_j})\over \Delta(A) \Delta(X)}.
\label{HCIZ}
\ee
\er
\br
The condition that the eigenvalues $\{a_i\}$ and $\{b_i\}$ of $A$ and $B$ be distinct can be eliminated
simply by taking limits in which some or all of these are made to coincide.  In the resulting determinantal formulae, like (\ref{CauchyBinet_Schur}), and those appearing in subsequent sections, in which a Vandermonde determinant $\Delta(A)$ or $\Delta(B)$ appears in the denominator, the only modification is that the terms in the numerator determinants depending on the $a_i$'s and $b_i$'s are replaced by their derivatives with respect to these parameters, taken to the same degree as the degeneracy of their values, while the denominator Vandermonde determinants are correspondingly replaced by their lower dimensional analogs. 
This will not be further developed here, but will be considered  elsewhere, in connection with correlation kernels for externally coupled matrix models. All formulae below in which no Vandermonde
determinant factors $\Delta(A)$ or $\Delta(B)$ appear in the denominator remain valid in the case of degenerate eigenvalues.
\er

\section{Applications to matrix models}

We now consider $N\times N$  matrix Hermitian  integrals that are  $\tau$-functions, and show how the application of convolution symmetries leads to new matrix models of the externally coupled type.
In the following, let $d\mu(M)$, be a measure on the space of $N\times N$
Hermitian matrices $M\in \Hb^{N\times  N}$ that is invariant under conjugation by  unitary matrices,
and such that the  reduced measure, projected to the space of eigenvalues by integration  over the  group $U(N)$,  is a product of $N$ identical measures $d\mu_0$  on $\Rb$, times the Jacobian factor $\Delta^2(X)$,
\be
\int_{U\in U(N)} d\mu(U X U^\dag) = \prod_{a=1}^N d\mu_0(x_a) \Delta^2(X).
\label{U(N)reduced_measure}
\ee
where $X = \diag(x_1, \dots , x_N)$.

\subsection{Convolution symmetries, externally coupled Hermitian matrix models
and $\tau$-functions as  finite determinants}

It is well-known that Hermitian matrix integrals of the form 
\bea
Z_N({\bf t}) &\&= \int_{M\in \Hb^{N \times N}} d\mu(M)
   \,    e^{\tr\sum_{i=1}^\infty t_i  M^i}\\
&\&= \prod_{a=1}^N\int_\Rb d\mu_0(x_a) e^{\sum_{i=1}^\infty t_i x_a^i} \Delta^2(X), 
\label{ZNdef_tau}
\eea
are KP-Toda $\tau$-functions \cite{ZKMMO}. The Schur function expansion is 
\be 
Z_N({\bf t})  = \sum_{ \ell(\lambda) \le N} \pi_{N,d\mu}(\lambda) s_\lambda({\bf t}),
\ee
where the coefficients $\pi_{N,d\mu}(\lambda)$ are expressible as determinants   in terms of the matrix of moments \cite{HO1, HO2, HO3} 
\bea
\pi_{N,d\mu}(\lambda) &\& \phantom{:} =   \prod_{a=1}^N\left(\int_\Rb d\mu_0(x_a)\right)
 \Delta^2(X) s_\lambda([X]) 
 \label{1matrix_plucker}
  \\
&\& \phantom{:}  = (-1)^{{1\over 2}N(N-1)}N! \ \det(\MM_{\lambda_i +N -i, j -1}) |_{1\le i,j \le N} \\
\MM_{ij}  &\&:= \int_\Rb d\mu_0(x)  x^{i+j}  .
 \eea

Now consider the externally coupled matrix model integral (cf. refs. \cite{BH, DW, ZJ1, ZJ2})
\be
Z_{N, ext}(A):=  \int_{M\in \Hb^{N\times  N}}d\mu(M) e^{\tr (AM)}, 
\label{ZNext}
\ee
where $A\in  \Hb^{N\times  N}$  is a fixed $N\times N$ Hermitian matrix. This can be obtained by simply applying a convolution symmetry transformation  of  the type given in {\bf Example \ref{example1}} to the $\tau$-function defined by the matrix integral  ({\ref{ZNdef_tau}).

\begin{proposition}
\label{convdefZNext_exp}
Applying the convolution symmetry $ \tilde{C}_\rho$  to  the $\tau$-function  $Z_N({\bf t})$, where $\rho_+(z)$ and $\rho_-(z)$ are  defined as in (\ref{expconv}), (\ref{identconv}), and choosing the KP flow parameters as ${\bf t} = \left[A\,\right]$  gives, within a multiplicative constant,  the externally coupled matrix  integral {\rm (\ref{ZNext}) }
   \be
 \tilde{C}_\rho (Z_N)( [A\,]) = \left(\prod_{i=1}^{N-1} i!\right) ^{-1}Z_{N, ext}(A) .
 \label{convZN_ZNext_exp}
 \ee

\end{proposition}
{\bf Proof:}
Substituting the expansion \cite{HO3}
\be
e^{\tr AM} = \sum_{\ell(\lambda)\le N}  {d_{\lambda, N} \over (N)_\lambda}  s_\lambda([AM]), 
\ee
into (\ref{ZNext}), where 
\be 
d_{\lambda, N}  =s_\lambda(\Ib_N)
\ee
is the dimension of the irreducible $GL(N)$ tensor representation of symmetry type $\lambda$, 
and expressing $M$ in diagonalized form as
\be
M = U X U^\dag, 
\ee
where
$U\in U(N)$ and $X= \diag(x_1,  \dots x_N)$, gives
\be
Z_{N, ext}(A)= \sum_{\ell(\lambda)\le N}  \int_{U\in U(N)}d\mu_{H}(U)  \prod_{a=1}^N\int_\Rb d\mu_0(x_a) e^{\sum_{i=1}^\infty t_i x_a^i} \Delta^2(X){d_{\lambda, N}\over (N)_\lambda}s_\lambda([AUXU^\dag]).
\ee
Evaluating the character integral  (\ref{haar_schur_int}) and  using (\ref{1matrix_plucker}), it follows that
\bea
Z_{N, ext}(A) &\&=  \sum_{\ell(\lambda)\le N} \prod_{a=1}^N\int_\Rb d\mu_0(x_a) e^{\sum_{i=1}^\infty t_i x_a^i} \Delta^2(X){1\over (N)_\lambda}s_\lambda([A]) s_\lambda([X]) \cr
&\& = \sum_{\ell(\lambda)\le N} {1\over (N)_\lambda} \pi_{N,d\mu}(\lambda) s_\lambda([A])\cr
&\&=\sum_{\ell(\lambda)\le N} (\prod_{i=1}^{N-1} i!) r_\lambda(N) s_\lambda([A]) \cr
&\&= (\prod_{i=1}^{N-1} i!) \tilde{C}_\rho (Z_N)\big\vert_{{\bf t} = [A\,]}.
\eea
where the third line follows from the expression (\ref{rlambdaN_ex1}) for $r_\lambda(N)$ in example \ref{example1} and the last from Proposition 
\ref{convdefZNext_exp}, eq. (\ref{conv_tau}). \hfill Q.E.D.

More generally, given an arbitrary function $\rho_+(z)$, analytic on the interior of $S^1$ and choosing 
$\rho_-(z)$ as in (\ref{identconv}), we may  define a new externally coupled matrix integral 
\be
Z_{N, \rho}(A) := \int_{M\in \Hb^{N \times N}} d\mu(M)\, \tau_r(N, [AM]), 
\label{ztaur_int}
\ee
in which  $e^{\tr AM}$ is replaced by\be
\tau_r(N, [M]) := \tau_r(N, [\Ib_N], [M])
= \sum_{\ell(\lambda) \le N} d_{\lambda, N} r_\lambda(N) s_\lambda([M]).
\ee
Then by the same calculation as above, it follows that $Z_{N, \rho}(A) $ is again just the $\tau$-function obtained by applying the convolution symmetry $\tilde{C}_\rho$ to $Z_N$, evaluated at the  parameter values ${\bf t} = [A\,]$.

\begin{proposition}
\label{convdefZNext}
Applying the convolution symmetry $\tilde{C}_\rho$  to $Z_N$ gives 
\be
\tilde{C}_\rho (Z_N)([A\,])
=Z_{N, \rho}(A).
\label{convZN_ZNext}
\ee
\end{proposition}
\smallskip \noindent
In particular, if we take $(\rho_+, \rho_-)$ as in {\bf Example \ref{example2}} above, we obtain (cf. \cite{HO3})
\be
Z_{N, \rho}(A)= \left( \prod_{i=0}^{N-1}  {(a)_i\over i!} \right) \zeta^{{1\over 2} N(N-1)} \int_{M\in \Hb^{N \times N}} d\mu(M) \, \det(1- \zeta A M)^{-a-N+1},
\ee
showing that this also is a KP-Toda $\tau$-function evaluated at parameter values ${\bf t} = [A\,]$.

Returning to the general case, a  finite determinantal  formula for $Z_{N, \rho}(A)$ is given by the following.

\begin{proposition}
\label{deterpZNrhoA}
\be
   Z_{N, \rho}(A) =
    {(-1)^{{1\over 2}N(N-1)}  N!\over \Delta(A)}  \det(G_{ij}(\rho, A))\big\vert_{1\le i,j \le N}, 
      \label{andreif_Delta_rho}
      \ee
    where
  \be
  G_{ij}(\rho, A) := \int_\Rb d\mu_0(x) x^{i-1}\rho_+(a_j x).
  \label{Gijdef}
    \ee
\end{proposition}
{\bf Proof:}
Applying the character integral identity (\ref{haar_schur_int}) to (\ref{ztaur_int}) gives
\bea
   Z_{N, \rho}(A) &\& = \int_{M\in \Hb^{N \times N}} d\mu(M)
   \,     \sum_{\ell(\lambda)\le N} r_\lambda(N) s_\lambda([A]) s_\lambda([M]) \\
    &\& =  {1\over \Delta(A)} \int d\mu_0(X)\Delta(X)  
       \det(\rho_+(a_i x_j))\big\vert_{1\le i,j \le N}
      \label{CauchyBinet_SchurAM}
        \\
      &\& = {(-1)^{{1\over 2}N(N-1)}  N!\over \Delta(A)}  \det(G_{ij}(\rho, A))\big\vert_{1\le i,j \le N}, 
  \label{detGij}
      \eea
    with $G_{ij} (\rho, A)$  defined by (\ref{Gijdef}). Here, the integration over the $U(N)$ group has been performed and {\bf Lemma \ref{dethypergeom}} has been used  in eq.~(\ref{CauchyBinet_SchurAM}). Eq.~(\ref{detGij}) follows from  (\ref{CauchyBinet_SchurAM}) by applying the Andr\'eief identity \cite{An} in the form
   \be
\left( \prod_{m=1}^N   \int d\mu_0(x_m) \right)\det(\phi_i(x_j))   \det(\psi_k (x_l))\Big\vert_{1\le i,j \le N \atop 1\le k,l \le N} 
   =N! \det\left ( \int \phi_i(x) \psi_j(x)\right)\Big\vert_{1\le i,j \le N}
   \ee
   with
   \be
   \phi_i(x) = x^{N-i}, \quad \psi_j(x) := \rho_+(a_j x),
   \ee
since
   \be
   \Delta(X) = \det(\phi_i(x_j)).
   \ee
   \hfill Q.E.D.

\subsection{Externally coupled two-matrix models}

We now turn to the case of two-matrix models. For simplicity, we only  consider the Itzykson-Zuber exponential coupling \cite{IZ}, although the same double convolution transformations may be applied to all the couplings considered in ref.~\cite{HO3}. Using the HCIZ identity (\ref{HCIZ})  to evaluate the integrals over the unitary groups $U(N)$, we obtain
\bea
Z_N^{(2)}({\bf t}, {\bf  \tilde t}) &\&= \int_{M_1 \in  \Hb^{N\times  N}} d\mu(M_1)\int_{M_2  \in \Hb^{N\times  N}}  d\tilde{\mu}(M_2) \ e^{\tr(\sum_{i=1}^\infty\left( t_i M_1^i +\tilde{ t}_i M_2^i)+ M_1M_2\right)} 
\label{ZN2red}\\
&\&= \prod_{k=1}^N k!\prod_{a=1}^N\left(\int_\Rb d\mu_0(x_a)\int_\Rb d\tilde{\mu}_0(y_a) 
\ e^{\sum_{i=1}^\infty( t_i x_a^i + \tilde{t}_i y_a^i + x_ay_a)} \right)\Delta(X)\Delta(Y),
\nonumber 
\eea
where $Y = \diag(y_1, \dots, y_N)$. This is known to be a 2KP-Toda $\tau$-function \cite{AvM1, AvM2, HO1, HO2, HO3, OS},  with  double Schur function expansion
\be
Z_N^{(2)}({\bf t}, {\bf  \tilde t}) =
\sum_\lambda \sum_\mu B_{N,d\mu, d\tilde{\mu}}(\lambda, \mu) s_\lambda ({\bf t})
s_\mu (\tilde{{\bf t}}),
\ee
where the coefficients $B_{N,d\mu, d\tilde{\mu}}(\lambda, \mu) $ are $N \times N$ determinants of submatrices  in terms of the matrix of bimoments
\bea
B_{N,d\mu, d\tilde{\mu}}(\lambda, \mu) &\&=
 \prod_{k=1}^N k!\prod_{a=1}^N\left(\int_\Rb d\mu_0(x_a) \int_\Rb d\tilde{\mu}_0(y_a) e^{x_ay_a}\right)
 \Delta(X)  \Delta(Y)s_\lambda([X]) s_\mu([Y])  \cr
 &\& \cr
&\& \phantom{:}  = (N! )\prod_{k=1}^N k!\, \det(\BB_{\lambda_i -i +N , \, \mu_j-j +N}) |_{1\le i,j \le N} \\
\BB_{i j}  &\&:=\int_\Rb d\mu_0(x_a) \int_\Rb d\tilde{\mu}_0(y_a) e^{x_ay_a}x^iy^j .
\eea

Now, choosing a pair of elements $(\rho, \tilde{\rho})$, with both $\rho_-$ and $\tilde{\rho}_-$  as in
 (\ref{identconv}), we may define a family of externally coupled two-matrix models, by
\be
Z^{(2)}_{N, \rho, \tilde{\rho}}(A,B) :=  \int_{M_1 \in  \Hb^{N\times  N}} \hspace{-24 pt}d\mu(M_1)\int_{M_2  \in \Hb^{N\times  N}}  \hspace{-24 pt} d\tilde{\mu}(M_2)\  \tau_r (N, [A], [M_1])  \tau_{\tilde{r}}(N, [B], [M_2]) e^{\tr( M_1M_2)}.
\label{ZN2extAB}
\ee
where  $A, B$ are a pair of hermitian $N \times N$ matrices.
This class may be obtained as the 2KP-Toda $\tau$-function resulting from applying the convolution symmetry $\tilde{C}_{\rho, \tilde{\rho}}$ to $Z_N^{(2)}$.

\begin{proposition}
\label{convZN2extAB}
Applying the convolution symmetry $\tilde{C}_{\rho, \tilde{\rho}}$ to $Z_N^{(2)}$ and evaluating at
the parameter values ${\bf t} = [A]$, ${\bf \tilde t} = [B]$ gives the externally coupled matrix integral 
(\ref{ZN2extAB})
\be
 \tilde{C}^{(2)}_{\rho, \tilde{\rho}}(Z_N^{(2)}) ( [A], [B]))
 =Z_{N, \rho, \tilde{\rho}}^{(2)}(A, B) .
\ee
\end{proposition}
{\bf Proof:}
Because of the $U(N) \times U(N)$ invariance of the measures $d\mu$ and $d\tilde{\mu}$ in  (\ref{ZN2extAB}) and all factors in the integrand, except for the coupling term $e^{\tr( M_1M_2)}$, we may carry out the two $U(N)$ integrations,  using  the HCIZ identity (\ref{HCIZ}), to obtain a reduced integral over the diagonal matrices $X = \diag(x_1, \dots , x_N)$ , $Y= \diag(y_1, \dots , y_N)$ of eigenvalues of $M_1$ and $M_2$,
\bea
Z_{N, \rho, \tilde{\rho}}^{(2)}(A, B) &\& =  
\prod_{k=1}^N k!\prod_{a=1}^N\left(\int_\Rb d\mu_0(x_a)\int_\Rb d\tilde{\mu}_0(y_a) 
\ e^{ x_ay_a} \right)\Delta(X)\Delta(Y) \\
&\& {\hskip 70 pt} \times  \tau_{C_\rho} (N, [A], [X])  \tau_{C_{\tilde{\rho}}}(N, [B], [Y]) 
\cr
&\& = \sum_{\ell(\lambda)\le N} \sum_{\ell(\mu)\le N}  r_\lambda(N)B_{N,d\mu, d\tilde{\mu}}(\lambda, \mu) 
\tilde{r}_\lambda(N)
s_\lambda ([A]) s_\mu ([B])\\
&\& = \tilde{C}^{(2)}_{\rho, \tilde{\rho}}(Z_N^{(2)}) ( [A], [B])).
\eea
where the second equality follows from eq. (\ref{hypergeom2KPtau}) and the last from Proposition \ref{conv_tau},
eq. (\ref{convtau2Ngschurexp}).   \hfill Q.E.D.

Since the dependence on $A$ and $B$ is $U(N) \times U(N)$ conjugation invariant we may choose, without loss of generality, $A$ and $B$ to be diagonal matrices 
\be
A=\diag(a_1, \dots , a_N),  \quad B= \diag(b_1, \dots , b_N),
\ee
We then obtain, as in the one-matrix case,  a  finite determinantal  formula for the 2KP-Toda $\tau$-function $Z^{(2)}_{N, \rho, \tilde{\rho}}(A, B)$.

\begin{proposition}
        \label{andreif_rho_tilderho}
\be
 Z_{N, \rho, \tilde{\rho}}^{(2)}(A, B) =
  {N!(\prod_{k=1}^N k! )\over \Delta(A)\Delta(B) } \det(G_{ij}(\rho, \tilde{\rho} , A,B)\big\vert_{1\le i,j \le N}, 
 \label{CauchyBinet_SchurABM2}
 \ee
 where
  \be
  G_{ij}(\rho, \tilde{\rho} , A,B) := \int_\Rb d\mu_0(x)\int_\Rb d\tilde{\mu}_0(y) e^{xy} 
  \rho_+(a_i x) \tilde{\rho}_+(b_j y).
    \ee
    \end{proposition}
 {\bf Proof :}
\bea
   Z_{N, \rho, \tilde{\rho}}^{(2)}(A, B) &\& = \int_{M_1\in \Hb^{N \times N}} d\mu(M_1)
    \int_{M_2\in \Hb^{N \times N}} d\tilde{\mu}(M_2) e^{\tr(M_1M_2)} \cr
&\& \quad  \times 
  \sum_{\ell(\lambda)\le N} r_\lambda(N) s_\lambda([A\,]) s_\lambda([M_1]) 
    \sum_{\ell(\mu)\le N} \tilde{r}_\mu(N) s_\mu([B\, ]) s_\mu([M_2]) \\
    &\& =  { (\prod_{k=1}^N k! ) 
     \over \Delta(A)\Delta(B)} \int d\mu(X)  \int d\tilde{\mu}(Y) 
    \, e^{\sum_{i=1}^N x_i y_i}
      \cr
    &\&   {\hskip 60 pt}    \times
       \det(\rho_+(a_k x_l))\big\vert_{1\le k,l \le N}
       \det(\tilde{\rho}_+(b_m y_n))\big\vert_{1\le m, n \le N}
      \label{CBSABM1}
        \\
      &\& = {N! (\prod_{k=1}^N k! )\over \Delta(A)\Delta(B)) } \det(G_{ij}(\rho, \tilde{\rho} , A,B)\big\vert_{1\le i,j \le N} .
      \label{CBSABM2}
      \eea
        In (\ref{CBSABM1}), we have used the  HCIZ identity  (\ref{HCIZ}),  antisymmetry of the determinants in the integrand with respect to permutations in the integration variables $(x_1, \dots , x_N)$ and $(y_1, \dots , y_N)$ and {\bf Lemma \ref{dethypergeom}} twice, while in (\ref{CBSABM2}), we have used the  Andr\'eief identity \cite{An} in the form
   \be
\left( \prod_{m=1}^N   \int d\mu(x_m, y_m) \right)\det(\phi_i(x_j))   \det(\psi_k (y_l))\Big\vert_{1\le i,j \le N \atop 1\le k,l \le N} 
   =N! \det\left ( \int d\mu(x,y)  \phi_i(x) \psi_j(y)\right)\Big\vert_{1\le i,j \le N}
   \label{andreief2}
   \ee
  \hfill Q.E.D.
 
  As the simplest example of a 2KP-Toda $\tau$-function obtained through {\bf Propositions    \ref{convZN2extAB}} and   {\bf \ref{andreif_rho_tilderho}}, consider the case when the measures  $d\mu_0(x)$ and $d\mu_0(y$) are both Gaussian, and $\rho_+$ and $\tilde{\rho}_+$  
  are both taken as the exponential function.
   \begin{example}
   {\rm \small
  \be 
  d\mu_0(x) = e^{-\sigma x^2}dx,  \quad  d\mu_0(y) = e^{-\sigma y^2}dy,
  \quad \rho_+(x) = e^x, \quad \tilde{\rho}_+(y) = e^y.
  \ee
  Evaluating the Gaussian integrals gives
  \be
  G_{ij} = {2\pi \over \sqrt{1+4 \sigma^2}} e^{\sigma(a_i^2 +  b_j^2) -a_ib_j
  \over 4\sigma^2-1}, 
  \ee
  and hence 
  \be
  Z_{N, \rho}(A)  =   { (2\pi)^N N!  \prod_{k=1}^N k !\over (1+4\sigma^2)^{N\over 2}\Delta(A) \Delta(B)}
    e^{{\sigma \over 4\sigma^2 -1}\sum_{i=1}^N (a_i^2 + b_i^2)} \det(e^{\sigma a_i b_j \over 1- 4\sigma^2}).
    \label{gaugedscaledHC}
  \ee
  The  factor $ e^{{\sigma \over 4\sigma^2 -1}\sum_{i=1}^N (a_i^2 + b_i^2)}$ is a linear exponential in terms of the 2KP flow variable $t_2$ and $\tilde{t}_2$  and hence, through the Sato formula
 (\ref{SatoBA}), produces just a gauge factor multiplying the Baker-Akhiezer function \cite{SW}. Therefore  (\ref{gaugedscaledHC}) is just a rescaled, gauge transformed version of the 2KP $\tau$-function of hypergeometric type appearing in the integrand of the Itzykson-Zuber coupled two-matrix model \cite{IZ}.
  }
   \end{example}
\subsection{More general 2KP-Toda $\tau$-functions as multiple integrals}

We may  extend the above results to more general   $2$KP-Toda $\tau$-functions expressed as multiple integrals and finite determinants. To begin with, the following multiple integral
\be
\tau^{(2)}_{d\mu}(N, {\bf t}, {\bf  \tilde t}) =\prod_{a=1}^N\left( \int_\Gamma  \int_{\tilde{\Gamma} }d\mu(x_a, y_a) 
e^{\sum_{i=1}^\infty( t_i x_a^i + \tilde{t}_i y_a^i)} \right)\Delta(X)\Delta(Y),
\label{2KPmult_int_tau}
\ee
where $\Gamma$, $\tilde{\Gamma}$ are curves in the complex $x$- and $y$-planes and $d\mu(x,y)$ is a measure on $\Gamma \times \tilde{\Gamma}$, is a 2KP-Toda $\tau$-function \cite{HO3} for a large class of measures $d\mu_0(x,y)$. Applying a  double convolution symmetry 
$\tilde{C}_{\rho, \tilde{\rho}}$,  with  $\rho_-$ and $\tilde{\rho}_-$  the same as in  (\ref{identconv}), gives a new 2KP-Toda $\tau$-function, also having a multiple integral representation.

\begin{proposition}
\label{conv2KP}
 \be
\tilde{C}^{(2)}_{\rho, \tilde{\rho}}(\tau^{(2)}_{d\mu})(N, {\bf t},{\bf \tilde t})= \prod_{a=1}^N\left( \int_\Gamma  \int_{\tilde{\Gamma} }d\mu(x_a, y_a)\right)
 \Delta(X)\Delta(Y)  \tau_{r} (N, {\bf t}, [X]) \tau_{\tilde{r}}(N, {\bf {\tilde t}}, [Y]) .
 \label{conv2KP_multiple_int}
\ee
\end{proposition}
{\bf Proof:}  This is proved similarly to {\bf Proposition \ref{convZN2extAB}}, using the Cauchy-Littlewood identity (\ref{tau2NId}) twice in the form
\be
\prod_{a=1}^Ne^{\sum_{i=1}^\infty( t_i x_a^i + \tilde{t}_i y_a^i)}  = \sum_{\ell(\lambda)\le N}s_\lambda({\bf t})
s_\lambda([X])  \sum_{\ell(\mu)\le N} s_\mu({\bf \tilde t}) s_\mu([Y]).
\ee
\hfill Q.E.D.
  
Evaluating at  parameter values ${\bf t}=[A]$ and ${\bf \tilde t}=[B]$, and applying again {\bf Lemma \ref{dethypergeom}} gives the $\tau$-function of eq.~(\ref{conv2KP_multiple_int}) in $N\times N$  determinantal form.

\begin{proposition}
\label{conv_multiple_int_det}
\be
\tilde{C}^{(2)}_{\rho, \tilde{\rho}}(\tau^{(2)}_{d\mu})([A], [B])={N!  \over \Delta(A) \Delta(B)}\det(G_{ij}(\rho, \tilde{\rho} , A,B)\big\vert_{1\le i,j \le N}, 
      \label{CauchyBinet_SchurABM3}
      \ee
    where
  \be
  G_{ij}(\rho, \tilde{\rho} , A,B) := \int_\Gamma \int_{\tilde{\Gamma}} d\mu(x,y)
  \rho_+(a_j x) \tilde{\rho}_+(b_j y).
        \label{andreif_dmu_rho_tilderho}
    \ee 
    \end{proposition}
    {\bf Proof:}
    \bea
\tilde{C}^{(2)}_{\rho, \tilde{\rho}}(\tau^{(2)}_{d\mu})([A], [B]) &\&=
{1\over \Delta(A) \Delta(B)} \prod_{a=1}^N\left( \int_\Gamma  \int_{\tilde{\Gamma} }d\mu(x_a, y_a)\right)    \cr
    &\&   {\hskip 70 pt}    \times
       \det(\rho_+(a_k x_l))\big\vert_{1\le k,l \le N}
       \det(\tilde{\rho}_+(b_m y_n))\big\vert_{1\le m, n \le N} \cr
   &\& = {N!  \over \Delta(A) \Delta(B)}\det(G_{ij}(\rho, \tilde{\rho} , A,B))\big\vert_{1\le i,j \le N}, 
       \eea
where again, we have used the {\bf Lemma \ref{dethypergeom}} twice and the Andr\'eief identity in the form (\ref{andreief2}).   
 \hfill Q.E.D.
   
  This therefore provides a new class of 2KP-Toda $\tau$-functions expressible in such a finite determinantal form, associated to any pair of curves $\Gamma$, $\tilde{\Gamma}$, together with a measure $d\mu$ on their product, and a pair of  functions  $\rho_+(x)$ and $\tilde{\rho}_+(y)$,  such that the integrals in  (\ref{andreif_dmu_rho_tilderho}) are well defined and convergent.

\bigskip \bigskip
\noindent
{\em Acknowledgements.} The authors would like to thank D.~Wang for helpful discussions relating to this work.  

\bigskip


\end{document}